\documentclass[oneside,onecolumn,12pt,a4paper]{article}
\usepackage[margin=20mm]{geometry}
\usepackage{graphicx}
\usepackage{dsfont}
\usepackage[round,authoryear]{natbib}
\usepackage{bbding}
\usepackage{setspace}
\usepackage{braket}
\usepackage{sectsty}
\usepackage{tikz}
\usepackage[justification=centering,margin=20mm]{caption}
\usepackage{amsmath}
\usepackage{mathtools}
\usepackage{amssymb}
\usepackage{hyperref}
\usepackage{color}
\definecolor{darkblue}{rgb}{0.0,0.0,0.3}
\hypersetup{colorlinks,breaklinks,linkcolor=darkblue,urlcolor=darkblue,anchorcolor=darkblue,citecolor=darkblue}

\def\ci{{\;\perp\!\!\!\perp\;}}

\def\citepos#1{\citeauthor{#1}'s (\citeyear{#1})}

\usepackage{ifthen}
\def\eprinttmp@#1arXiv:#2 [#3]#4@{\ifthenelse{\equal{#3}{}}{\href{http://arxiv.org/abs/#1}{arXiv:#1}}{\href{http://arxiv.org/abs/#2}{arXiv:#2 [#3]}}}
\newcommand{\eprint}[1]{\eprinttmp@#1arXiv: []@}
\newcommand{\doi}[1]{\href{http://dx.doi.org/#1}{doi:#1}}

\sectionfont{\normalfont\normalsize\bfseries}
\subsectionfont{\normalfont\small\bfseries}
\subsubsectionfont{\normalfont\small\bfseries}

\makeatletter
\g@addto@macro\quote{\small\singlespacing\upshape\sffamily\vspace{-4mm}}
\makeatother

\begin{document}

\sloppy

\setstretch{1.2}

\title{A sideways look at faithfulness for quantum correlations}
	\author{Peter W. Evans}
	\maketitle

\begin{abstract}
  Despite attempts to apply the lessons of causal modelling to the observed correlations typical of entangled bipartite quantum systems, Wood and Spekkens argue that any causal model purporting to explain these correlations must be fine tuned; that is, it must violate the assumption of faithfulness. The faithfulness assumption is a principle of parsimony, and the intuition behind it is basic and compelling: when no statistical correlation exists between the occurrences of a pair of events, we have no reason for supposing there to be a causal connection between them. This paper is an attempt to undermine the reasonableness of the assumption of faithfulness in the quantum context. Employing a symmetry relation between an entangled bipartite quantum system and a `sideways' quantum system consisting of a single photon passing sequentially through two polarisers, I argue that Wood and Spekkens' analysis applies equally to this sideways system. If this is correct, then the consequence endorsed by Wood and Spekkens for an ordinary entangled quantum system amounts to a rejection of a causal explanation in the sideways, single photon system, too. Unless rejecting this causal explanation can be sufficiently justified, then it looks as though the sideways system is fine tuned, and so a violation of faithfulness in the ordinary entangled system may be more tolerable than first thought. Thus extending the classical `no fine-tuning' principle of parsimony to the quantum realm may well be too hasty.
\end{abstract}

\section{Introduction: EPRB and faithfulness}
\label{sec:intro}

Towards the end of the twentieth century progress in the field of machine learning led to the development of algorithms that could automate the discovery of causes. These so-called causal discovery algorithms \citep{Spirtes00,Pearl09} permit an inference from given statistical dependences and independences between distinct measurable elements of some system to a causal model for that system. As part of the algorithms, a series of constraints must be placed on the resulting models that capture general features that we take to be characteristic of causation. There are two significant assumptions. The first is the causal Markov condition, which ensures that every statistical dependence in the data results in a causal dependence in the model---essentially a formalisation of Reichenbach's common cause principle. The second assumption, faithfulness, is that any resultant causal model faithfully reproduces the statistical dependences and independences; in other words, every statistical independence implies a causal independence (or, no causal independence is the result of a fine-tuning of the model).

Trouble arises when this causal discovery framework is employed to model certain quantum phenomena, particularly the observed Einstein-Podolsky-Rosen-Bell (EPRB) correlations in an entangled bipartite quantum system. To illustrate this, consider such an EPRB experiment, where each of the parts of the bipartite state are subject to a freely and independently specifiable local measurement, $\alpha$ and $\beta$, with two respective possible results for each measurement, $A$ and $B$. Figure~\ref{fig:eprb} shows such a system consisting of a pair of photons each passing through a polariser. The local measurement outcomes in an EPRB experiment on a maximally entangled state are correlated such that the joint probability ${\mathrm{P}(\mathrm{\textbf{A}} = \mathrm{\textbf{B}})}$ of the same result at each measurement device is given by ${\cos^{2}(\alpha -\beta)}$. Moreover, observed statistical conditional independences, ${(A \ci \beta \mid \alpha)}$ and ${(B \ci \alpha \mid \beta)}$,\footnote{$(X \ci Y \mid Z)$ denotes that $X$ and $Y$ are \emph{conditionally independent} given $Z$:
\begin{equation*}
  \mathrm{P}(X,Y \mid Z) = \mathrm{P}(X \mid Z)\cdot\mathrm{P}(Y \mid Z).
\end{equation*}} represent the fact that there can be no signalling from one side of the experiment to the other.

\begin{figure}[t]
	\begin{center}
		\begin{tikzpicture}[scale=1.8]
			\footnotesize
			\draw[color=black,->] (2.5,0)--(2.5,1) node [midway,sloped,anchor=center,above] {Time}; 
			\draw[color=gray,ultra thick] (1.75,1.25)--(1.25,1.75)--(1.75,2.25)--(2.25,1.75)--(1.75,1.25)--(1.25,1.75) node [left,black,text width=12mm] {Setting angle $\beta$}; 
			\draw[color=gray,thick] (1.75,1.4)--(1.75,2.1); 
			\draw[color=blue,densely dashed] (2,2)--(2.5,2.5) node [above,black] {$\mathrm{\textbf{B}} = 1$}; 
			\draw[color=blue,densely dashed] (1.5,2)--(1.0,2.5) node [above,black] {$\mathrm{\textbf{B}} = 0$}; 
			\draw[color=gray,ultra thick] (-1.75,1.25)--(-1.25,1.75)--(-1.75,2.25)--(-2.25,1.75)--(-1.75,1.25)--(-1.25,1.75) node [right,black,text width=12mm] {Setting angle $\alpha$}; 
			\draw[color=gray,thick] (-1.75,1.4)--(-1.75,2.1); 
			\draw[color=blue,densely dashed] (-1.5,2)--(-1,2.5) node [above,black] {$\mathrm{\textbf{A}} = 0$}; 
			\draw[color=blue,densely dashed] (-2,2)--(-2.5,2.5) node [above,black] {$\mathrm{\textbf{A}} = 1$}; 
			\draw[color=blue,thick] (0,0)--(1.5,1.5) node [midway,sloped,anchor=center,above,black] {Photon}; 
			\draw[color=blue,thick] (0,0)--(-1.5,1.5) node [midway,sloped,anchor=center,above,black] {Photon}; 
			\draw (0,0) node (O) [below] {Source};
			\draw (0,-0.5) node {$\mathrm{P}(\mathrm{\textbf{A}} = \mathrm{\textbf{B}}) = \cos^{2}(\alpha -\beta)$};
		\end{tikzpicture}
	\end{center}
	\caption{EPRB experiment with a pair of photons. (Adapted from (\citealp[p.7753]{PriceWharton15}; \citealp[p.302]{EvansPriceWharton}).)}
	\label{fig:eprb}
\end{figure}

\citet{WoodSpekkens} argue that any causal model purporting to explain the observed EPRB correlations must be fine tuned. By `fine tuned' they mean that what they call the `causal-statistical parameters', which ``specify a conditional probability distribution for every variable given its causal parents, $P(X \mid \text{Pa}(X))$'' \citep[p.4]{WoodSpekkens}, are precisely balanced so as to hide any conditional dependence between putatively causally dependent variables.\footnote{Such parameters are `specified' during the process of model building in the context of causal modelling. For a successful causal model, this specification will be constrained by the observed (in)dependences.} According to the causal modelling framework, the faithfulness assumption states that every statistical independence implies a causal independence. Applied to the EPRB scenario, since the observed statistical independences in such a system imply no signalling between the parties---that is, a statistical independence---one must infer that there can be no (direct or mediated) causal link from one side of the experiment to the other. However, the joint probability over the outcomes, $\mathrm{\textbf{A}}$ and $\mathrm{\textbf{B}}$, indicates that there is a statistical dependence between them. According to faithfulness, though, we are unable to account for this dependence with a causal link unless this link is fine tuned to ensure that ${(A \ci \beta \mid \alpha)}$ and ${(B \ci \alpha \mid \beta)}$ still hold. There is thus a fundamental tension between the observed quantum correlations and the no-signalling requirement, the faithfulness assumption, and the possibility of a causal explanation.

More precisely, \citet[p. 24]{WoodSpekkens} show that the following three assumptions form an inconsistent set:
\begin{enumerate}
  \item The predictions of quantum theory are correct---that is, the conditional independences ${(A \ci \beta \mid \alpha)}$ and ${(B \ci \alpha \mid \beta)}$ are satisfied, and a Bell inequality is violated;
  \item The observed statistical dependences and independences can be given a causal explanation as per the causal discovery framework;
  \item The faithfulness assumption holds---that is, there is no fine-tuning.
\end{enumerate}
Wood and Spekkens conclude that, since the faithfulness assumption is an indispensable element of causal discovery, the second assumption must yield. The contrapositive of this is that \emph{any} purported causal explanation of the observed no-signalling EPRB correlations in an entangled bipartite quantum system falls afoul of the tension between the no-signalling constraint and `no fine-tuning' and, thus, must violate the assumption of faithfulness. Such causal explanations, so the argument goes, including retrocausal explanations, should therefore be ruled out as viable explanations.

What is it about the faithfulness assumption that would make it indispensable? The intuition behind the assumption is basic and compelling. When no statistical correlation exists between the occurrences of a pair of events, we have no reason for supposing there to be a causal connection between them. Conversely, if we were to allow the possibility of a causal connection between statistically uncorrelated events, we would have a particularly hard task determining which of these uncorrelated sets could be harbouring a conspiratorial causal connection that hides the correlation. We can thus think of the faithfulness assumption as a bit like Occam's razor: the simplest explanation for a pair of statistically uncorrelated events is that they are causally independent. \citet[p. 48]{Pearl09} illustrates this point nicely with an example of a picture of a chair: we do not expect (i.e. it is unlikely for) a picture of a chair to be a picture of two chairs perfectly aligned such that one hides the other.

There are well-known examples of systems that potentially show a misapplication of the faithfulness assumption. One such example, originating in \citet{Hesslow}, involves a contraceptive pill that can cause thrombosis while simultaneously lowering the chance of pregnancy, which can also cause thrombosis. As \citet[p. 246]{Cartwright01} points out, given the right weights for these processes, it is conceivable that the net effect of the pills on the frequency of thrombosis be zero. This is a case of `cancelling paths', where the effect of two or more causal routes between a pair of variables cancels to achieve statistical independence. In a case such as this, since we can have independent knowledge of the separate causal mechanisms involved here, we have grounds for arguing that there really is a causal connection between the variables despite their statistical independence. Thus, we can certainly imagine a scenario in which the faithfulness assumption could lead us astray. However, in defence of the general principle, an example such as this clearly contains what Wood and Spekkens refer to as fine-tuning; the specific weights for these processes would need to match precisely to erase the statistical dependence, and such a balance we would generally think of as unstable (any change in background conditions, \emph{etc}. would reveal the causal connection in the form of a statistical dependence).

This paper is an attempt to make trouble for the assumption of faithfulness in the quantum setting.\footnote{\citet{Naeger16} proposes a related explanation for the unfaithfulness of EPRB correlations. His proposal is that unfaithfulness is unproblematic so long as it occurs in a stable way, such that any change in background conditions maintains the unfaithful independences. He suggests that the fine-tuning mechanism in quantum mechanics is what he calls `internal cancelling paths', and is analogous to the ordinary cancelling paths scenario just considered. For a detailed critique of N\"{a}ger's proposal, and its relation to the sort of retrocausal models considered below, see \citet{Evans2018}. Evans argues that the source of unfaithfulness in a basic retrocausal model can be interpreted as an example of N\"{a}ger's internal cancelling paths.} The focus of this trouble is a very simple quantum system investigated in \citet{Price12}, \citet*{EvansPriceWharton}, and \citet{PriceWharton15}, consisting of a single photon passing sequentially through two polarisers. This single photon quantum system is noteworthy because it can be related on symmetry grounds \citep{EvansPriceWharton,LeiferPusey17} to the entangled bipartite quantum system from Figure~\ref{fig:eprb}. Since the single photon system is a temporally oriented version of the spatial correlations of the EPRB system, \citet[p.302]{EvansPriceWharton} refer to it as `sideways' EPRB, or SEPRB. Given that certain control constraints on the initial input of the photon in the sideways system can be formulated to be the \emph{temporal reverse} of the output of the ordinary system, I argue here that Wood and Spekkens' inconsistent set of assumptions can be applied to this sideways system also.

If this analysis is correct, three options present themselves: (i) the sideways system is as resistant to causal explanation as the ordinary entangled system; (ii) a violation of faithfulness in the ordinary entangled system is as tolerable as it is in the sideways system; or (iii) the symmetries relating the ordinary and sideways systems carry no weight in guiding our response to Wood and Spekkens' inconsistent set. I contend that the strength of the symmetry relation rules out option (iii). Option (ii) provides grounds for accepting the presence of fine-tuning to explain the observed correlations in an entangled bipartite quantum system. If this option is to be rejected, as Wood and Spekkens argue, then a compelling case must be made for option (i). In the absence of such a case, extending the classical `no fine-tuning' principle of parsimony to the quantum realm may well be too hasty. In so far as `no fine-tuning' is an impediment to the possibility of local hidden variables,\footnote{Of the sort that could underpin the $\psi$-epistemic approach to quantum mechanics.} abandoning local hidden variables on account of the `no fine-tuning' principle may well be too hasty also.

\section{How signalling and causation come apart}

Our interest in this paper is with an experimental set-up closely related to the EPRB set-up of Figure~\ref{fig:eprb} (in \S\ref{sec:symmetry} I will argue that this relation is very close indeed). There are two modifications to the EPRB set-up that can be made to obtain in the new experiment. The first modification (proposed by \citet[p.7756]{PriceWharton15}) is to `reflect' one half of the experiment in time, so to speak, in order to produce a set-up consisting of a single photon that passes through two polarisers sequentially. This is the `sideways' system, SEPRB. The second modification (proposed by \citet[p.77]{Price12}) relates to the nature of what now have become input channels on the earlier polariser. In the interests of temporal symmetry, Price introduces a demon\footnote{This demon has been referred to in conference talks by Price as the ``demon of the left'', and as ``Erutan'' in \citep[p.7757]{PriceWharton15}.} that controls the inputs to the system to ensure that these are the time-symmetric representation of the outputs. This second modification leads us to what we will call `input-controlled' SEPRB.

\begin{figure}[t]
	\begin{center}
		\begin{tikzpicture}[scale=1.8]
			\footnotesize
			\draw[color=black, ->] (2.5,-2)--(2.5,1) node [midway,sloped,anchor=center,above] {Time}; 
			\draw[color=gray,ultra thick] (1.75,1.25)--(1.25,1.75)--(1.75,2.25)--(2.25,1.75)--(1.75,1.25)--(1.25,1.75) node [left,black,text width=12mm] {Setting angle $\beta$}; 
			\draw[color=gray,thick] (1.75,1.4)--(1.75,2.1); 
			\draw[color=blue,densely dashed] (2,2)--(2.5,2.5) node [above,black] {$\mathrm{\textbf{B}} = 1$}; 
			\draw[color=blue,densely dashed] (1.5,2)--(1.0,2.5) node [above,black] {$\mathrm{\textbf{B}} = 0$}; 
			\draw[color=gray,ultra thick] (-1.75,-1.25)--(-1.25,-1.75)--(-1.75,-2.25)--(-2.25,-1.75)--(-1.75,-1.25)--(-1.25,-1.75) node [right,black,text width=12mm] {Setting angle $\alpha$}; 
			\draw[color=gray,thick] (-1.75,-1.4)--(-1.75,-2.1); 
			\draw[color=blue,densely dashed] (-1.5,-2)--(-1,-2.5) node [below,black] {$\mathrm{\textbf{A}}^{\prime} = 0$}; 
			\draw[color=blue,densely dashed] (-2,-2)--(-2.5,-2.5) node [below,black] {$\mathrm{\textbf{A}}^{\prime} = 1$}; 
			\draw[color=blue,thick] (-1.5,-1.5)--(1.5,1.5) node [midway,sloped,anchor=center,above,black] {Photon}; 
		\end{tikzpicture}
	\end{center}
	\caption{SEPRB experiment with a single photon. (Adapted from \citep[p.7757]{PriceWharton15}.)}
	\label{fig:seprb}
\end{figure}

The resulting experimental set-up is shown in Figure~\ref{fig:seprb}. The first thing to note about the SEPRB set-up is that the correlations between $\mathrm{\textbf{A}}^{\prime}$ and $\mathrm{\textbf{B}}$ are exactly the same as the correlations between $\mathrm{\textbf{A}}$ and $\mathrm{\textbf{B}}$ in EPRB; that is, ${\mathrm{P}(\mathrm{\textbf{A}}^{\prime} = \mathrm{\textbf{B}}) = \cos^{2}(\alpha -\beta)}$. Thus, the observed statistical correlations between $\mathrm{\textbf{A}}^{\prime}$ and $\mathrm{\textbf{B}}$ will violate a timelike analogue of Bell's local causality condition, and so a CHSH inequality \citep{LeiferPusey17}.\footnote{A further demonstration of the power of this analogy between the ordinary spacelike version of the CHSH inequalities for EPRB and their timelike analogue is demonstrated by \citet{Henaut18}, who show that the timelike version maintains a game-theoretic quantum advantage over classical systems, despite no possibility of nonlocality or contextuality.} The second thing to note is the nature of the input channels at $A^{\prime}$ as a result of the second modification. The input at $A^{\prime}$ for input-controlled SEPRB is the temporal reverse of the output at $A$ for EPRB---that is, the output at $A$ for EPRB cannot be chosen by the experimenter so, likewise, the input at $A^{\prime}$ for SEPRB cannot be chosen either, rather we imagine it to be independently and randomly generated (by a demon, say). Due to the discrete nature of photons, a single photon must enter the input polariser exclusively from either the left, ${\mathrm{\textbf{A}}^{\prime} = 1}$, or the right, ${\mathrm{\textbf{A}}^{\prime} = 0}$.\footnote{Another justification for this feature of the sideways experiment that does not rely on assuming photon discreteness would be simply to employ time symmetry between the outputs and inputs as per \citep{LeiferPusey17}.} An experimenter setting the angle $\alpha$ at $A^{\prime}$ is clearly making a difference to the polarisation of the photon that exits the polariser, albeit this control extends only so far as the photon polarisation up to an additive factor of $\frac{\pi}{2}$: the photon could have a polarisation $\alpha$ if the input came from the left, or a polarisation ${\alpha + \frac{\pi}{2}}$ if the input came from the right. But due to the fact that the direction from which the input photon arrives is randomly generated, there is no possibility to use that difference making to send a signal to the experimenter at $B$. That is, regardless of the input channel and setting $\alpha$, ${\mathrm{P}(\mathrm{\textbf{B}} = 0) = \mathrm{P}(\mathrm{\textbf{B}} = 1) = \frac{1}{2}}$.

This is straightforward to see. Firstly, it is instructive to note that without the `input control', there is no natural possibility for violations of faithfulness---that is, the causal relation between the setting $\alpha$ and the subsequent state of the photon manifests in the conditional probability distributions over the outcomes at $B$. Since the correlations between $\mathrm{\textbf{A}}^{\prime}$ and $\mathrm{\textbf{B}}$ are ${\mathrm{P}(\mathrm{\textbf{A}}^{\prime} = \mathrm{\textbf{B}}) = \cos^{2}(\alpha - \beta)}$ and ${\mathrm{P}(\mathrm{\textbf{A}}^{\prime} \neq \mathrm{\textbf{B}}) = \sin^{2}(\alpha - \beta)}$, the probability for a particular outcome at $B$ is the sum ${p\cos^{2}(\alpha - \beta) + (1 - p)\frac{1}{2}\sin^{2}(\alpha - \beta)}$, where $p \in [0,1]$ provides weights over the inputs at $A^{\prime}$. When we do not know the value of $\mathrm{\textbf{A}}^{\prime}$ (as is the case when this input is hidden and random) then $p = \frac{1}{2}$ and the probability of a particular outcome at $B$ becomes a sum of half the contribution from the ${\mathrm{\textbf{A}}^{\prime} = 0}$ channel and half the contribution from the ${\mathrm{\textbf{A}}^{\prime} = 1}$ channel, or ${\frac{1}{2}\cos^{2}(\alpha - \beta) + \frac{1}{2}\sin^{2}(\alpha - \beta) = \frac{1}{2}}$. As a result of this, it should be obvious that the probability of either output at $B$, given some setting $\beta$, is independent of the setting $\alpha$, ${(B \ci \alpha \mid \beta)}$. It should also be relatively uncontroversial to see that ${(A^{\prime} \ci \beta \mid \alpha)}$.

In this version of input-controlled SEPRB, then, we have a system that has the same correlations as EPRB, and so can violate a Bell inequality, and the same conditional independences as EPRB, and so satisfies the first condition of Wood and Spekkens' inconsistent set. This implies that any causal explanation of the observed statistical dependences must be fine tuned. Since the intuitive interpretation of SEPRB is that there is a straightforward causal explanation of the correlations (that the photon carries information about $\alpha$ from $A^{\prime}$ to $B$), we have a simple example of a system that violates faithfulness. But it does not seem that there is anything pernicious about this violation---in fact we can pinpoint exactly where the fine-tuning emanates in input-controlled SEPRB.

The face of fine-tuning in the sort of operational description that Wood and Spekkens prefer is a disconnect between signalling and causation. But in input-controlled SEPRB, this disconnect is grounded in the hidden randomness of the input. If the experimenter controlling $\alpha$ at $A^{\prime}$ were to discover the actual input channels for each photon, or somehow knew beforehand the pattern of inputs, it would be straightforward for the experimenter to use the polariser to control precisely the polarisation of each photon and send a signal. Without this knowledge---that is, when we have a hidden random input---no signal can be sent. This exposes the patent fact that the independence ${(B \ci \alpha \mid \beta)}$ only holds under a certain specification of conditional probability distributions (which are themselves constrained to reproduce the observed independences). Without this specific balance of the conditional probability distribution ${\mathrm{P}(B \mid \alpha) = \mathrm{P}(B) = \frac{1}{2}}$, which in this case is a straightforward reflection of the hidden randomness of the input $\alpha$, $B$ would be dependent upon $\alpha$ as we would intuitively expect from this sort of experimental arrangement. This balancing of the causal-statistical parameters to manufacture a statistical independence between causally dependent variables is precisely what we define as fine-tuning. Thus we can see then that this hidden randomness is a condition that, in the right quantum system, suffices to provide a violation of faithfulness.

We have here a system that violates faithfulness in a way that is completely explicable in terms of a specific epistemic constraint. We also have some obvious similarities between EPRB and SEPRB, not least that we have generated SEPRB through a kind of temporal reflection of one half of EPRB. It would be interesting to investigate the possibility that the same sort of explication for the source of fine-tuning in SEPRB is available for EPRB also. Conversely, the same similarities between EPRB and SEPRB might be used as the basis for arguing that a rejection of fine-tuning, and so a rejection of a causal explanation, in EPRB strengthens the case that fine-tuning and causal explanation should be rejected for SEPRB also. I will compose these options into a trilemma in \S\ref{sec:trilemma} (similar to the trilemma found in \citep{EvansPriceWharton}). But, of course, all of this hinges on the strength of the similarity between EPRB and input-controlled SEPRB. In the next section I provide two independent symmetry arguments in favour of a strong relation between EPRB and SEPRB.

\section{Symmetry arguments}
\label{sec:symmetry}

Before we consider the two symmetry arguments relating EPRB and SEPRB, let us briefly consider here to what the analogous source of fine-tuning for EPRB would amount, given our account above. The key point regarding the source of fine-tuning in SEPRB is that the value of the input $A^\prime$ is not available to be conditionalised upon for the purposes of establishing the relevant conditional probability distributions---that is, the causal-statistical parameters---relating the variables in the causal model. It is straightforward to see that this is also the case for EPRB, since by construction the outcome $\mathrm{\textbf{A}}$ is not relevant to the choice of measurement setting $\alpha$ (which is unsurprising, because the input of SEPRB was explicitly constructed to temporally mirror the left output of EPRB). But for clarity we can consider the analogous causal dependence that is obscured by fine-tuning.\footnote{For more detail on this analogy and the associated causal dependence, see \citep{PriceWharton15}.}

Consider the rather counterintuitive scenario that we can somehow know the outcome $\mathrm{\textbf{A}}$ of the measurement $A$ in EPRB before we make our choice of the value of the measurement setting $\alpha$. Since ${\mathrm{P}(\mathrm{\textbf{A}} = \mathrm{\textbf{B}}) = \cos^{2}(\alpha - \beta)}$, if we know that, say, ${\mathrm{\textbf{A}} = 0}$ as a result of whatever choice we make for $\alpha$, then, for a fixed choice $\beta$, the probability distribution over the values of $B$ would be dependent upon $\alpha$, i.e. ${\mathrm{P}(B \mid \alpha A) \neq \mathrm{P}(B)}$. This highlights the reason that this is a strange scenario to consider: what would be happening in such a situation is that our choice $\alpha$, to maintain outcome ${\mathrm{\textbf{A}} = 0}$, would seem to be influencing the polarisation of the incoming photon, and we do not usually consider our measurement outputs fixed and measurement inputs variable in such scenarios. Nevertheless, this (albeit unusual) dependence of $B$ on $\alpha$ is a dependence that is obscured by the particular conditional probability distributions over the outcomes at $A$ and $B$ that we customarily take to model the observed statistical dependences and independences in EPRB. On this account, the fact that this `no-signalling' independence is not typically exposed as a causal dependence is because we do not ordinarily conditionalise on the output variable when determining the other relevant conditional probability distributions, due to the natural epistemic constraint provided by our own temporal orientation with respect to the exogenous variables.

However, the two symmetry arguments that we will now consider lend support to the idea that we should take the analogy between EPRB and SEPRB seriously, and so also the counterintuitive causal structure. As such, in the background of these arguments sit a range of further arguments in favour of a retrocausal interpretation of the EPRB correlations. While addressing these further arguments is beyond the scope of the current proposal, see \citet{Price12,EvansPriceWharton,PriceWharton15} and references therein for further detail.

\subsection{Action symmetry}

\citet{EvansPriceWharton} employ what they call an `action symmetry' to argue in favour of like ontological ascriptions (or causal explanations, even) to EPRB and SEPRB.\footnote{The action symmetry is developed further in \citep{WhartonMillerPrice}.} They note the following deep similarity between the two systems \citep[p.306]{EvansPriceWharton}:
\begin{quote}
  they span bounded regions of spacetime with precisely the same electromagnetic action $S$. It should therefore be no surprise that the experimental correlations in EPRB and SEPRB are identical, as our most advanced theory of these interactions -- quantum electrodynamics (QED) -- reduces the joint probability to a functional integral of the classical action\ldots [under the right conditions] there is an exact ``action symmetry'' (S-symmetry) between these two experiments for any given outcome\ldots
\end{quote}

The upshot of this action symmetry is that one is able effectively to permute the polariser at $A^\prime$ from Figure~\ref{fig:seprb} to other spatiotemporal locations in an action-preserving way; in particular, one can permute the polariser at $A^{\prime}$ into the polariser at $A$ in Figure~\ref{fig:eprb} maintaining the same action. Thus we have a perfectly natural explanation for why the joint probabilities ${\mathrm{P}(\mathrm{\textbf{A}} = \mathrm{\textbf{B}})}$ and ${\mathrm{P}(\mathrm{\textbf{A}}^{\prime} = \mathrm{\textbf{B}})}$ are the same: ``they result from the same QED mathematics'' \citep[p.307]{EvansPriceWharton}.

Action-symmetry provides our first argument that there is more than just a superficial similarity between ERPB and SEPRB. As a result, we should expect equivalent tension in the two cases between any causal explanation and the faithfulness assumption. The next symmetry we consider here, operational symmetry, can be characterised as a more general symmetry principle that in a sense encompasses action-symmetry and provides even stronger grounds for taking the similarity between EPRB and SEPRB seriously.

\subsection{Operational symmetry}

In his analysis of quantum contextuality, \citet[p.052108-1]{Spekkens05} proposes a definition of a noncontextual ontological model in terms of operational equivalences:
\begin{quote}
  A noncontextual ontological model of an operational theory is one wherein if two experimental procedures are operationally equivalent, then they have equivalent representations in the ontological model.
\end{quote}
This definition amounts to something akin to Leibniz' Law for dealing with operational equivalences: no ontological difference without operational difference. Following \citet{Spekkens05} and, indeed, \citet{WoodSpekkens}, an operational model is a specification of a set of possible preparations, transformations, and measurements, along with associated conditional probability distributions for the relevant variables. Thus, operational equivalence is simply an equivalence of conditional probability distributions between corresponding variables. Significantly, conditional probability distributions do not necessitate a temporal direction, and thus it need not matter in which temporal direction the specific variables are related.

\citet[p.4]{LeiferPusey17}, in their response to \citet{Price12}, extend Spekkens' operational equivalence principle to motivate their assumption of time symmetry: ``a symmetry of the operational predictions ought to also hold at the ontological level''. Since for the case of EPRB and input-controlled SEPRB we have just such a symmetry of the operational predictions---a \emph{temporal} symmetry by construction that renders the outputs at $A$ operationally equivalent to the inputs at $A^{\prime}$---we should expect that any causal explanation (i.e. at the ontological level) should be equivalent between the two cases. And so we should expect equivalent tension in the two cases between any causal explanation and the faithfulness assumption.

Somewhat more speculatively, it appears that this operational symmetry might come in a stronger flavour. \citet{ShrapnelCosta2018} develop a generalised ontological models framework that does not include any assumptions about causal structure. Within their framework, one separates a system of interest into local laboratories with local controllables, and distinguishes between all those parts of the system that are correlated with the controllables, and the rest of the invariant structure (including whatever causal structure) of the environment that mediates the correlations between local observables (what they call the \emph{process}). With one small caveat that the local laboratory at $A^{\prime}$ is the temporal inverse of the local laboratory at $A$, EPRB and SEPRB are both members of the same equivalence class of processes. Applying Spekkens' operational equivalence principle amounts to claiming that there should be no ontological difference between EPRB and SEPRB, and so no difference in causal explanation. Thus, again, we should expect equivalent tension in the two cases between any causal explanation and the faithfulness assumption.

\section{A trilemma}
\label{sec:trilemma}

If these symmetry arguments are on the right track, then we can confidently say that whatever consequences arise for a causal explanation in EPRB should arise also for a causal explanation in SEPRB, and \emph{vice versa}. For instance, when we hold fixed the input at $A$ in EPRB (as is necessitated), our choice $\alpha$ influences the probability distribution over the value of the outcome $\mathrm{\textbf{A}}$, and so we ascribe a causal explanation to the correlation between our choice $\alpha$ and the output at $A$. Likewise, when we hold fixed the output at $A^{\prime}$ in SEPRB (albeit counterintuitively), it becomes clear that our choice $\alpha$ influences the probability distribution over the value of the input $\mathrm{\textbf{A}}^{\prime}$ (by constraining, say, the choice that the demon can make to ensure the fixed output at $A^{\prime}$), and so the above symmetries dictate that in this counterintuitive scenario we should similarly ascribe a causal explanation to the correlation between our choice $\alpha$ and the input at $A^{\prime}$.\footnote{We do not usually think of this sort of causal relation as one that we can exploit for purposes of control since, due to our inherent temporal orientation, we are never in a position to hold fixed outputs in this way; we usually take the input at $A^{\prime}$ to be exogenous, and the output endogenous, not the other way around.}

We are now in a position to present the trilemma I alluded to in \S\ref{sec:intro}.
\begin{enumerate}
	\item \label{enum:i} SEPRB is as resistant to causal explanation as EPRB on account of rejecting fine-tuning; 
	\item \label{enum:ii} a violation of faithfulness in EPRB is as tolerable as it is in SEPRB;
	\item \label{enum:iii} the symmetries relating EPRB and SEPRB carry no weight in guiding our response to Wood and Spekkens' analysis.
\end{enumerate}

Given the strength of the arguments in the previous section, I set aside option (\ref{enum:iii}) rejecting the symmetries, but note that this does remain a live option. This leaves options (\ref{enum:i}) and (\ref{enum:ii}), each of which by my lights stands or falls on the potency of the narrative I have presented for input-controlled SEPRB. I wish to finish off this analysis with some general discussion points that bear upon how one might approach the residual dilemma.

The first point to make concerns the origin of the fine-tuning in SEPRB. The fine-tuning we find here comes from the randomisation of the inputs at $A^{\prime}$, independent of and hidden from the local experimenters. This inaccessible randomisation of inputs is motivated by time symmetry to be the temporal reverse of the operational process that characterises the stochastic nature of the outputs at $A$. But when viewed in the context of SEPRB, our intuition for a causal explanation is overwhelmingly strong and, once we are aware of the process by which the `demon' feeds the input channels, such a causal explanation does not seem objectionable or mysterious. Moreover, understanding that the demonic process is the temporal reverse of an ordinary stochastic process further ameliorates any concern about any sort of cosmic conspiracy aligning the causal parameters of the system to erase the possibility of signalling---`fine-tuning' in this sense is just what we would expect to arise from this specific type of epistemic constraint.

\citet{LeiferPusey17} explicitly point out, in their response to \citepos{Price12} discussion of what we have called here input-controlled SEPRB, that some sort of `no fine-tuning' assumption must be violated. They outline that they take the most legitimate way of dealing with fine-tuning to be by accounting for the fine-tuning as some emergent feature of the system. They indicate that perhaps the absence of signalling to the past and its uncoupling with any potential retrocausal influence could arise from the same process from which the thermodynamic arrow emerges, so explaining the fine-tuning without basing it on a fundamental physical principle. The current proposal is consistent with this account of emergent fine-tuning: a violation of faithfulness in SEPRB is by construction a function of the limited epistemic access an experimenter has to the input channels. But the purpose of this construction was to emulate the temporal inverse of the randomised outputs from a polarising beam splitter in the EPRB experiment, where the outputs arise stochastically. So the analogy in the EPRB case would be that the fine-tuning required by a causal explanation for the correlations in the face of no-signalling emerges from the randomness of the outputs, rather than being a fundamental feature of any physical system.

Let us now consider if we were to reject fine-tuning, as Wood and Spekkens do, what option (\ref{enum:i}) would entail. Pursuing option (\ref{enum:i}) requires us not only to reject any (retro)causal explanation of the correlations between $A$ and $B$ in EPRB, but also to reject the causal narrative that we tell to explain why the input at $A^{\prime}$ is correlated with the output at $B$ in SEPRB. In the latter case we ordinarily find it straightforward to explain the correlation in an SEPRB experiment between the inputs and outputs: the classical causal narrative claims that the photon carries with it information about its polarisation state upon which the outcome is conditional. If rejecting fine-tuning means rejecting classical causal explanations such as this, then we would no longer be able to avail ourselves of this intuitive picture underlying the statistical correlations between input and output. Let us not underestimate the importance of such causal explanation---beyond simply accounting for observed phenomena, a causal explanation provides \emph{understanding} of a physical process. In the absence of a causal explanation, we lack insight into the unobservable world. Relatedly, in so far as this sort of classical causal picture is necessary for a realist ontology, option (\ref{enum:i}) would be pushing us towards a necessary operationalism about quantum phenomena.\footnote{Alternatively, one might provide an account of quantum causal explanation that explains both the EPRB and SEPRB experiments in terms of a quantum causal model \citep{CostaShrapnel16}.} While there is nothing necessarily wrong with rejecting classical causal explanation in this way, it is not a cost-free manoeuvre, and a more compelling case must be mounted in its favour.

\section{Concluding remarks}

Wood and Spekkens claim that the faithfulness assumption is an indispensable element of causal discovery. That may be correct---the specific machine learning algorithms that isolate causal structure may well require a constrained framework within which to operate. But this framework is motived by the relative scarcity of formal tools available for characterising causal reasoning in statistics, medicine, economics, social science, and especially the fields of artificial intelligence and cognitive science (\citet[p.xiii]{Pearl09}; \citet[p.xi]{PearlGlymourJewell16}). Causal discovery algorithms, and the causal assumptions that underpin them, come with no guarantee that they will be applicable to the observed correlations between and within quantum systems. This paper presents a case to undermine the reasonableness of the assumption of faithfulness in the quantum context. Short of a rejection of a causal explanation in a straightforward single photon system, I contend that this `sideways' system is, according to Wood and Spekkens' analysis, fine tuned. This at least partially mitigates the concern that entangled bipartite quantum systems themselves violate faithfulness. Thus, extending the classical `no fine-tuning' principle of parsimony to the quantum realm may well be too hasty. In so far as `no fine-tuning' is an impediment to the possibility of local hidden variables, abandoning local hidden variables on account of the `no fine-tuning' principle may well be too hasty also.

As a final note of warning, even if a violation of faithfulness in the EPRB experiment might be given a `natural' explanation along the lines given above, and so diminish such violation as a sticking point for retrocausal explanations of the correlations between entangled quantum systems, a much more serious challenge faces any explanation of quantum phenomena in terms of exotic causal structures. \citet{ShrapnelCosta2018} argue that any non-contextual ontological model incorporating exotic causal structure cannot match the observed statistical predictions of quantum mechanics. That is, ontological models accounting for observed quantum statistics must be necessarily contextual. But in so far as contextuality is a kind of fine-tuning \citep{Cavalcanti18}, the sort of `natural' explanation of violations of faithfulness in SEPRB looks like exactly what would be needed to provide a justification of a contextual ontological model employing exotic causal structure explaining quantum statistics. Whether this can be done, however, is an open problem.

\end{document}